# The Speed of Light and the Einstein Legacy: 1905-2005

Reginald T. Cahill
School of Chemistry, Physics and Earth Sciences,
Flinders University, GPO Box 2100, Adelaide 5001, Australia
Reg.Cahill@flinders.edu.au

## Abstract

That the speed of light is always c (≈300,000 km/s) relative to any observer in nonaccelerating motion is one of the foundational concepts of physics. Experimentally this was supposed to have been first revealed by the 1887 Michelson-Morley experiment, and was made one of Einstein's key postulates of *Special Relativity* in 1905. However in 2002 the actual 1887 fringe shift data was analysed for the first time with a theory for the Michelson interferometer that used both the Fitzgerald-Lorentz contraction effect, as well as the effect of the air on the speed of light. That analysis showed that the data gave an absolute motion speed in excess of 300 km/s. So far six other experiments have been shown to give the same result. This implies that the foundations of physics require significant revision. As well data shows that both Newtonian gravity and General Relativity are also seriously flawed, and a new theory of gravity is shown to explain various so-called gravitational `anomalies', including the `dark matter' effect. So the centenary of Einstein's *Special Relativity* turns out to be also its demise. Most importantly absolute motion is now understood to be the cause of the various relativistic effects, in complete contradiction with the Einstein viewpoint, but in accord with the earlier proposal by Lorentz.

## Introduction

Physicists have believed for more than 100 years that the speed of light was the same for all nonaccelerating observers. This is one of the postulates that Einstein proposed in 1905, with the first experimental evidence supposed to have been based upon the results of the Michelson-Morley 1887 experiment. However a re-analysis in 2002 of the fringe-shift data from that experiment showed that the data indicated a speed in excess of 300 km/s. So far another six experiments have been found that also revealed a similar speed. It means that space has a substructure, that absolute motion is observable, that the spacetime construct has no ontological meaning, despite its use in quantum field theory, and that it is absolute motion that causes the various relativistic effects, as first suggested by Lorentz. These discoveries have profound implications for the foundations of physics and for our understanding of reality. Einstein asserted that that there is no preferred frame, that there is no detectable *space*, that a three-dimensional *space* has no physical existence. However all through the last 100 years and more the experimental data was indicating that this assertion was invalid. Here a brief review of that data is presented along with another recent discovery, namely that our understanding of gravity was flawed from the very beginning, for it is now clear that Newtonian gravity is valid only in special situations, and that the Einstein theory of gravity, known as General Relativity, inherited these flaws, as discussed herein.

The Einstein postulates were first formulated in 1905 and have played a fundamental role in limiting the form of subsequent physical theories, and in also defining our comprehension of reality. They lead to the concept of spacetime, and that a curved spacetime explained gravity. They also lead physicists to reject any evidence that was revealing that the postulates were in disagreement with experimental data. In physics they have become a vigorously defended belief system, and any discussion of the numerous experiments that indicate their failure is banned.

**Einstein postulates:**
*(1) The laws of physics have the same form in all inertial reference frames.*
*(2) Light propagates through empty space with a definite speed c independent of the speed of the observer (or source).*
*(3) In the limit of low speeds the gravity formalism should agree with Newtonian gravity.*

The putative successes of the postulates lead to the almost universal acceptance of the Einstein *Special Theory of Relativity*, which is based upon the concept of a flat four-dimensional spacetime ontology that replaces the older separate concepts of space and time, and then to the *General Theory of Relativity* with its curved spacetime model for gravity. While the relativistic effects are well established experimentally it is was dramatically understood in 2002 [4,10] that numerous experiments, beginning with the Michelson-Morley experiment [1] of 1887, have always shown that postulates (1) and (2) (excepting the 2nd part) are false, namely that there is a detectable physical local frame of reference or `space', and that the solar system has a large observed galactic velocity of some 420±30km/s in the direction (RA=5.2hr, Dec= -67deg) through this space [2,3,5,8,10]. This is different from the speed of 369km/s in the direction (RA=11.20hr, Dec= -7.22deg) extracted from the Cosmic Microwave Background (CMB) anisotropy, and which describes a motion relative to the distant universe, but not relative to the local space. This paper briefly reviews the experimental evidence regarding the failure of the postulates, and the implications for our understanding of fundamental physics, and in particular for our understanding of gravity. A new theory of gravity is seen to be necessary, and this results in an explanation of the `dark matter' effect, entailing the discovery that the fine structure constant is a 2nd gravitational constant [2-4,6,7]. This theory is a part of the information-theoretic modelling of reality known as *Process Physics* [2-4,9], which premises a non-geometric process model of time, as distinct from the current *non-Process Physics,* which is characterised by a geometrical model of time.

## Detecting Absolute Motion in 1887

The first detection of absolute motion, that is motion relative to space itself, was actually by Michelson and Morley in 1887. However they totally bungled the reporting of their own data, an achievement that Michelson managed again and again throughout his life-long search for experimental evidence of absolute motion. The Michelson interferometer was a brilliantly conceived instrument for the detection of absolute motion, but only in 2002 was its principle of operation finally understood and used to analyse, for the first time ever, the data from the 1887 experiment, despite the enormous impact of that experiment on the foundations of physics, particularly as they were laid down, erroneously as we now understand, by Einstein. So great was Einstein's influence that the 1887 data was never re-analysed *post*-1905 using a proper relativistic-effects based theory for the interferometer. For that reason modern-day vacuum Michelson interferometer experiments are badly conceived, and their null results continue to cause much confusion: only a Michelson interferometer in the gas-mode can detect absolute motion, as we now see.

The Michelson interferometer, Fig.1, compares the change in the difference between travel times, when the device is rotated, for two coherent beams of light that travel in orthogonal directions between mirrors; the changing time difference being indicated by the shift of the interference fringes during the rotation. This effect is caused by the absolute motion of the device through space with speed $v$. The post relativistic-effects theory for this device is remarkably simple. The Fitzgerald–Lorentz contraction effect causes the arm AB parallel to the absolute velocity to be *physically* contracted to length

$$L_\| = L\sqrt{1 - \frac{v^2}{c^2}}$$

The time to travel AB is set by $Vt_{AB} = L_\| + vt_{AB}$, while for BA by $Vt_{BA} = L_\| - vt_{BA}$, where $V = c/n$ is the speed of light, with $n$ the refractive index of the gas present (we ignore here the Fresnel drag effect for simplicity – an effect caused by the gas also being in absolute motion). For the total ABA travel time we then obtain

$$t_{ABA} = t_{AB} + t_{BA} = \frac{2LV}{V^2 - v^2}\sqrt{1 - \frac{v^2}{c^2}}$$

For travel in the AC direction we have from the Pythagoras theorem $(Vt_{AC})^2 = L^2 + (vt_{AC})^2$ and that $t_{CA} = t_{AC}$. Then for the total ACA travel time

$$t_{ACA} = t_{AC} + t_{CA} = \frac{2L}{\sqrt{V^2 - v^2}}$$

Then the difference in travel time is

$$\Delta t = \frac{(n^2 - 1)L}{c} \frac{v^2}{c^2} + O(\frac{v^4}{c^4}) \qquad (1)$$

after expanding in powers of $v/c$. This clearly shows that the interferometer can only operate as a detector of absolute motion when not in vacuum ($n = 1$), namely when the light passes through a gas, as in the early experiments (in solids a more complex phenomenon occurs and rotation fringe shifts from absolute motion do not occur). A more general analysis [2,3], including Fresnel drag, gives

$$\Delta t = k^2 \frac{L v_P^2}{c^2} \cos(2(\theta - \psi)) \qquad (2)$$

where $k^2 \approx n(n^2 - 1)$, while neglect of the Fitzgerald-Lorentz contraction effect gives $k^2 = n^3 \approx 1$ for gases, which is essentially the Newtonian theory that Michelson used. The interferometers are operated with the arms horizontal, as shown by Miller's interferometer in Fig.2. Then in (2) θ is the azimuth of one arm relative to the local meridian, while ψ is the azimuth of the absolute motion velocity projected onto the plane of the interferometer, with projected component $v_P$. Here the Fitzgerald-Lorentz contraction is a real dynamical effect of absolute motion, unlike the Einstein spacetime view that it is merely a mathematical artefact, and whose magnitude depends on the choice of observer.

The Michelson and Morley air-mode interferometer fringe shift data revealed a speed of some 8km/s [1,4,10] when analysed using the prevailing but incorrect Newtonian theory which has $k^2 \approx 1$ in (2); and this value was known to Michelson and Morley. The often-repeated statement that Michelson and Morley did not see any rotation induced fringe shifts is completely wrong; all physicists should read their paper [1] for a re-education. They rejected their own data on the sole but spurious ground that the detected 8km/s was smaller than the speed of the earth about the sun of 30km/s. What their result really showed was that (i) absolute motion had been detected, and (ii) that the theory giving $k^2 \approx 1$ was wrong. Michelson and Morley in 1887 should have announced that the speed of light did depend of the direction of travel, that the speed $c$ was relative to an actual physical space. It was Miller [5] who saw the flaw in the 1887 paper and, while not having a theory for the value of $k^2$, he used the effect, throughout a year in 1925/26 and making 20,000 rotations of the interferometer at Mt. Wilson, of the changing vector addition of the earth's orbital velocity and the absolute galactic velocity of the solar system to determine the numerical value of $k^2$, and determined the first estimate for the absolute linear velocity of the solar system.

Including the Fitzgerald-Lorentz dynamical contraction effect as well as the effect of the gas present we find that $k^2 = 0.00058$ for air, which explains why the observed fringe shifts were so small. In the Einstein theory $k = 0$; absolute motion is to be undetectable in principle. Fig.3 shows data from averaging the fringe shifts from 20 rotations of the Miller interferometer, performed over a short period of time, and clearly shows the expected form in (2) (only a linear drift caused by temperature effects on the arm lengths has been removed). In Fig.3 the fringe shifts during rotation are given as fractions of a wavelength, $\Delta\lambda/\lambda = \Delta t/T$, where $\Delta t$ is given by (3) and $T$ is the period of the light. Such rotation induced fringe shifts clearly show that the speed of light is different in different directions. The claim that Michelson interferometers, operating in gas-mode, do not produce fringe shifts under rotation is clearly nonsense. But it is that claim that lead to the continuing belief, within physics, that absolute motion had never been detected, and that the speed of light is invariant. The value of $\psi$ from such rotations together lead to plots like those in Fig.4, which show $\psi$ from the 1925/1926 Miller [5] interferometer data for four different months of the year, from which the RA = 5.2hr is readily apparent. While the orbital motion of the earth about the sun slightly affects the RA in each month, and Miller used this effect do determine the value of $k$, the new theory of gravity required a reanalysis of the data [2-4,8], giving the above absolute velocity. Two old interferometer experiments, by Illingworth (1927) and Joos (1930), used helium, enabling the refractive index effect to be recently confirmed. The data from an interferometer experiment by Jaseja *et al* (1964) using two orthogonal masers also indicates that they detected absolute motion, but were not aware of that as they used the incorrect Newtonian theory and so considered the fringe shifts to be too small to be real, reminiscent of the same mistake by Michelson and Morley

Fig.5 shows the NS orientated coaxial cable RF travel time variations measured by DeWitte in 1991, which gives the same RA [2-4,8]. That experiment showed that RF EM waves travel at speeds determined by the orientation of the cable relative to the Miller direction; in complete contradiction to the Einstein postulate. That these very different experiments show the same speed and RA of absolute motion is one of the most startling but suppressed discoveries of the twentieth century. Torr and Kolen (1981) using an EW orientated nitrogen gas-filled coaxial cable also detected absolute motion. It should be noted that analogous optical fibre experiments give null results for the same reason, apparently, that transparent solids in a Michelson interferometer also give null results.

So Postulates (1) and (2) are in disagreement with the experimental data. In all some seven experiments have detected this absolute motion. Modern interferometer experiments use vacuum with $n = 1$, and then from (2) $k = 0$, predicting no rotation-induced fringe shifts. In analysing the data from these experiments the consequent *null* effect is misinterpreted to imply the absence of absolute motion. As discussed in [2-4] it is absolute motion which causes the dynamical effects of length contractions, time dilations and other relativistic effects, in accord with Lorentzian interpretation of relativistic effects.

## A New Theory of Gravity and the Fine Structure Constant

We now come to postulate (3) for gravity. This postulate relates General Relativity to Newtonian gravity, and Newtonian gravity is now known to be seriously flawed, and so *ipso facto,* by using this postulate, Einstein and Hilbert inadvertently developed a flawed theory of gravity. Newtonian gravity was based upon Kepler's Laws for the planetary motions within the solar system and uses the acceleration field $g(r,t)$,

$$\nabla \cdot g = -4\pi G\rho \qquad (3)$$

where $G$ is Newton's universal gravitational constant, and $\rho$ is the density of matter. However equally valid mathematically is a velocity field formulation [2,3,7,8]

$$\frac{\partial}{\partial t}(\nabla \cdot v) + \nabla \cdot ((v \cdot \nabla)v) = -4\pi G\rho \qquad (4)$$

with $g$ now given by the Euler fluid total derivative, which has the convective acceleration as its 2nd component

$$g = \frac{dv}{dt} \equiv \frac{\partial v}{\partial t} + (v \cdot \nabla)v \qquad (5)$$

External to a spherical mass $M$ a static velocity-field solution of (4) is

$$v(r) = -\sqrt{\frac{2GM}{r}}\hat{r} \qquad (6)$$

which gives from (5) the usual inverse square law

$$g(r) = -\frac{GM}{r^2} \qquad (7)$$

The velocity field formalism (4)-(5) is mathematically equivalent to (3); because they give the same acceleration field. However it is the velocity field formalism that turns out to be more fundamental because (i) it is observable, and (ii) it cannot be determined from the observed acceleration field. However (4), unlike (3), is not uniquely determined by Kepler's laws because

$$\frac{\partial}{\partial t}(\nabla \cdot v) + \nabla \cdot ((v \cdot \nabla)v) + C(v) = -4\pi G\rho \qquad (8)$$

where

$$C(v) = \frac{\alpha}{8}((trD)^2 - tr(D^2)) \qquad (9)$$

and

$$D_{ij} = \frac{1}{2}\left(\frac{\partial v_i}{\partial x_j} + \frac{\partial v_j}{\partial x_i}\right) \qquad (10)$$

also has the same external solution (6), as $C(v) = 0$ for the flow in (6). So the presence of the $C(v)$ dynamics would not have manifested in the special case of planets in orbit about the massive central sun. Here α is a dimensionless constant - a new additional gravitational constant. However inside a spherical mass $C(v) \neq 0$ and using the Greenland ice-shelf bore hole $g$ anomaly data, Fig.6, we find [6] that $\alpha^{-1} = 139 \pm 5$, which gives the fine structure constant $\alpha = 1/137$ to within experimental error. Here the gravity residual from (7) and (8) is $\Delta g(d) = -2\pi\alpha G\rho d$, where $\rho$ is the density of the ice and $d$ is the depth.

From (8) and (9) we can write the $C(v)$ term as an additional effective `matter density' on the RHS of (4)

$$\rho_{DM} = \frac{\alpha}{32\pi G}((trD)^2 - tr(D^2)) \qquad (11)$$

as a phenomenological treatment of the new spatial dynamics within the velocity field formulation of Newtonian gravity. It is this spatial dynamics that has been misinterpreted as the `dark matter' effect. However the dynamics in (8) cannot be written in terms of a closed form like (3) for the acceleration field; the `dark-matter' effect is only describable in a closed form using the velocity field formalism. So from the very beginning the Newtonian theory of gravity missed a major dynamical aspect of the phenomenon of gravity. This `dark matter' dynamical effect also appears to explain the long-standing problems in measuring $G$ in Cavendish-type experiments, as shown in Fig.7.

Eqn.(8) has novel black hole spherically symmetric and static solutions [6] where the radial in-flow magnitude is

$$v(r) = \left(\frac{K}{r} + \beta\left(\frac{1}{r}\right)^{\frac{\alpha}{2}}\right)^{1/2} \qquad (12)$$

where the key feature is the $\alpha$-dependent term in addition to the usual `Newtonian' in-flow in (6). For the in-flow in (12) the centripetal acceleration relation $v_o = \sqrt{rg(r)}$, with $g(r)$ given by (5) and (12), gives for circular orbits orbital rotation speeds of the form

$$v_o(r) = \frac{1}{2}\left(\frac{K}{r} + \frac{\beta\alpha}{2}\left(\frac{1}{r}\right)^{\frac{\alpha}{2}}\right)^{1/2} \qquad (13)$$

which is characterised by their almost flat asymptotic limit. This rotation curve explains the `dark matter' effect as seen in spiral galaxies, as shown in Fig.8. Such black holes can exist independent of matter ($K = 0$), and are not formed by an in-fall of matter (as are black holes in GR); they can be primordial, that is, formed during the big bang. Such black holes act as the kernels of spiral galaxies which are formed, early in the history of the universe, by subsequent rapidly in-falling matter, with that in-fall resulting in the rotating disk. Non-rotating elliptical galaxies do not form about primordial black holes, but are formed by normal Newtonian gravitational attraction between the matter. The spatial in-flow into these systems then results in a non-primordial black hole(s) forming. Globular clusters also exhibit such non-primordial black holes [6].

Eqns.(4) and (8) are only applicable to a zero vorticity flow. More generally vorticity is a relativistic effect (involving the speed of light $c$) and is given by

$$\nabla \times (\nabla \times v) = \frac{8\pi G \rho}{c^2} v_R \qquad (14)$$

where $v_R$ is the absolute velocity of the matter relative to the local space; see [3,6,11] for the more general form of (8) that includes vorticity and other relativistic effects. Fig.9 shows the component of the vorticity field $\nabla \times v$ induced by the rotation of the earth, while Fig.10 shows the much larger component induced by the earth's absolute linear motion. Eqn.(14) explains the Lense-Thirring `frame-dragging' effect in terms of vorticity in the flow field, but makes predictions different from General Relativity because of the absolute linear motion induced component. These conflicting predictions will soon be tested by the Gravity Probe B [14,15] gyroscope precession satellite experiment. Vorticity is the rotation of local space relative to more distant space, and this rotation is detectable by observing the spin direction of a gyroscope being simply carried around by the rotating space, the direction remaining fixed relative to the local space. The smaller component of the frame-dragging or vorticity effect caused by the earths absolute rotation component of $v_R$ has already been determined from the laser-ranged satellites LAGEOS(NASA) and LAGEOS2(NASA-ASI) [16] and the data implies the indicated coefficient on the RHS of (14) to ±10%. However that experiment cannot detect the larger component of the `frame-dragging' or vorticity induced by the absolute linear motion component of the earth as that effect is not cumulative, while the rotation induced component is cumulative. The predicted GP-B gyroscope spin precessions caused by the earth's absolute linear motion are shown in Fig.11.

Both (4) and (8) have wavelike aspects to their time-dependent solutions, with the time-dependence being the rule rather than the exception [2,3,7,12]. Such wave behaviour has been detected in most absolute motion experiments, as seen in the DeWitte data in Fig.5. For (4) these waves do not produce a gravitational force effect via (5), as seen by noting that (4) gives the same $g$ field as (3). But with the inclusion of the `dark matter' spatial dynamics in (8) such waves do produce a gravitational force effect, and there are various experimental `anomalies' which are probably manifestations of this effect. As well such waves affect the vorticity from (14), and so, in principle, could be detected by the GP-B experiment [12]. General Relativity predicts a very different kind of gravitational wave, but these have never been seen, despite extensive searches. The new theory of gravity implies that these waves do not exist.

The trajectory of test particles in the differentially flowing space are determined by extremising the proper time

$$\tau[r_o] = \int dt \left(1 - \frac{v_R^2}{c^2}\right)^{1/2} \qquad (15)$$

where $v_R = v - v_O$, with $v_O$ the velocity of the object, and $r_o$ its position, relative to an observer frame of reference, and $v$ is the velocity of space relative to the same frame of reference, which gives from (15) an acceleration independent of the test mass, in accord with the equivalence principle,

$$\frac{dv_O}{dt} = \left(\frac{\partial v}{\partial t} + (v.\nabla)v\right) + (\nabla \times v) \times v_R - \frac{v_R}{1 - \frac{v_R^2}{c^2}} \frac{1}{2} \frac{d}{dt}\left(\frac{v_R^2}{c^2}\right) \qquad (16)$$

Here the 1st term is the Euler `fluid' acceleration in (5), the 2nd term is the vorticity-induced Helmholtz acceleration, and the last is a relativistic effect leading to the so-called `geodesic' effects, such as the precession of elliptical orbits. It is significant that the time-dilation effect in (15) leads to the `fluid' acceleration in (4) and (5), revealing a close link between the spatial flow phenomena and relativistic effects. Indeed the form of the Euler `fluid' derivative in (5) and (16) is seen to be directly responsible for Newton's inverse square law of gravity. Reformulation of (15) for electromagnetic waves gives the gravitational light bending and in particular the gravitational lensing effect in place of (16).

## General Relativity and the Schwarzschild Metric

We saw that Newtonian gravity eventually failed because it was expressed in the limited formalism of the gravitational acceleration field $g$. As soon as we introduce the velocity field formalism together with its `dark matter' generalisation we see that numerous gravitational anomalies are explained [6]. General Relativity was constructed to agree with Newtonian gravity, and so it is flawed by this connection. So it is interesting to understand why General Relativity (GR) is supposed to have passed key observational and experimental tests. GR uses the Einstein tensor $G_{\mu\nu}$ and relates it to the energy-momentum stress tensor $T_{\mu\nu}$. Then (17) determines the metric $g_{\mu\nu}(x)$ tensor, where $d\tau^2 = g_{\mu\nu}dx^\mu dx^\nu$ specifies the proper time spacetime interval,

$$G_{\mu\nu} \equiv R_{\mu\nu} - \frac{1}{2}Rg_{\mu\nu} = \frac{8\pi G}{c^2}T_{\mu\nu} \qquad (17)$$

In this formalism the trajectories of test objects are also determined by extremising (15) which, after a general change of coordinates, gives the acceleration in (16) in terms of the usual affine connection $\Gamma^\lambda_{\mu\nu}$,

$$\Gamma^\lambda_{\mu\nu}\frac{dx^\mu}{d\tau}\frac{dx^\nu}{d\tau} + \frac{d^2x^\lambda}{d\tau^2} = 0 \qquad (18)$$

In the case of a spherically symmetric mass $M$ the well known solution of (17) outside of that mass is the external Schwarzschild metric

$$d\tau^2 = \left(1 - \frac{2GM}{c^2r}\right)dt^2 - \frac{r^2}{c^2}\left(d\theta^2 + \sin^2(\theta)d\varphi^2\right) - \frac{dr^2}{c^2\left(1 - \frac{2GM}{c^2r}\right)} \qquad (19)$$

This solution is the basis of various experimental checks of General Relativity in which the spherically symmetric mass is either the sun or the earth. The four tests are: the gravitational redshift, the bending of light, the precession of the perihelion of Mercury, and the time delay of radar signals. However the solution (19) is in fact completely equivalent to the in-flow interpretation of Newtonian gravity. Making the change of variables, first discovered by Panlevé and Gullstrand in the 1920's, $t \rightarrow t'$ and $r \rightarrow r' = r$ with

$$t' = t + \frac{2}{c}\sqrt{\frac{2GMr}{c^2}} - \frac{4GM}{c^2}\tanh^{-1}\sqrt{\frac{2GM}{c^2 r}} \tag{20}$$

the Schwarzschild solution (19) takes the form

$$d\tau^2 = dt'^2 - \frac{1}{c^2}\left(dr' + \sqrt{\frac{2GM}{r'}}dt'\right)^2 - \frac{r'^2}{c^2}\left(d\theta'^2 + \sin^2(\theta')d\varphi'^2\right) \tag{21}$$

which is exactly the differential form of (15) for the velocity field given by the Newtonian form in (6). This result shows that the Schwarzschild metric in GR is completely equivalent to Newton's inverse square law: GR in this case is nothing more than Newtonian gravity in disguise. So the so-called `tests' of GR were nothing more than a test of the `geodesic' equation (15), where most simply this is seen to determine the motion of an object relative to an observable and observed absolute local physical 3-space. These tests were merely confirming the in-flow formalism, and have nothing to do with a Schwarzschild spacetime ontology. So the claimed `successful tests' of GR had nothing to do with the conjectured GR Einstein dynamics; and as well whenever GR was found to fail, as for the spiral galaxy non-Keplerian rotations, for example, spurious arguments were put in place to protect the theory, here by introducing the concept of `dark matter' and the consequent costly but fruitless search for such `matter', or in numerous other cases the experimental results were suppressed.

Since GR has only been directly tested using the metric in (19) or equivalently (21), it is interesting to ask what is the particular form that (17) then takes. To that end we substitute into (17) a special class of flow-metrics – the Panlevé-Gullstrand metric, involving an arbitrary time-dependent velocity flow-field,

$$d\tau^2 = g_{\mu\nu}dx^\mu dx^\nu = dt^2 - \frac{1}{c^2}(dr - v(r,t)dt)^2 \tag{22}$$

The various components of the Einstein tensor are then found to be [2,3,7]

$$G_{00} = \sum v_i \overline{G}_{ij} v_j - c^2 \sum \overline{G}_{0j} v_j - c^2 \sum v_i \overline{G}_{i0} + c^2 \overline{G}_{00}$$
$$G_{i0} = -\sum \overline{G}_{ij} v_j + c^2 \overline{G}_{i0}$$
$$G_{ij} = \overline{G}_{ij} \tag{23}$$

where the $\overline{G}_{\mu\nu}$ are given by

$$\overline{G}_{00} = \frac{1}{2}((trD)^2 - tr(D^2))$$
$$\overline{G}_{i0} = \overline{G}_{0i} = -\frac{1}{2}(\nabla \times (\nabla \times v))_i$$
$$\overline{G}_{ij} = \frac{d}{dt}(D_{ij} - \delta_{ij} trD) + \left(D_{ij} - \frac{1}{2}\delta_{ij} trD\right)trD - \frac{1}{2}\delta_{ij} tr(D^2) - (D\Omega - \Omega D)_{ij} \tag{24}$$

and where $\Omega_{ij} = \frac{1}{2}\left(\frac{\partial v_i}{\partial x_j} - \frac{\partial v_j}{\partial x_i}\right)$ is the tensor form of the flow vorticity, and most interesting here $d/dt$ is the Euler fluid total derivative. In vacuum, with $T_{\mu\nu} = 0$, we see that $G_{\mu\nu} = 0$ implies that $\overline{G}_{\mu\nu} = 0$. The first equation in (24) then demands that

$$\rho_{DM} = \frac{\alpha}{32\pi G}((trD)^2 - tr(D^2)) = 0 \qquad (25)$$

This simply corresponds to the fact that GR does not permit the `dark matter' effect, and this happens because GR was forced to agree with Newtonian gravity, in the appropriate limits, and that theory also has no such effect. As well in GR the energy-momentum tensor $T_{\mu\nu}$ is not permitted to make any reference to absolute linear motion of the matter; only the relative motion of matter or absolute rotational motion is permitted, again another legacy of the erroneous Einstein postulates. It is very significant to note that the above exposition of the GR formalism for the metrics in (22) is exact. Taking the trace of $\overline{G}_{ij}$ in (24) we obtain using (10), also exactly, and in the case of zero vorticity and outside of matter, that

$$\frac{\partial}{\partial t}(\nabla.v) + \nabla.((v.\nabla)v) = 0 \qquad (26)$$

which is exactly the `velocity field' formulation in (4) of Newtonian gravity outside of matter. This should have been expected as it corresponds to the previous observation that the `Newtonian in-flow' velocity field is exactly equivalent to the external Schwarzschild metric. There is in fact only *one* indirect confirmation of the GR formalism apart from the misleading external Schwarzschild metric cases, namely the observed decay of the binary pulsar orbital motions, for only in this case is the metric non-Schwarzschild, and so not equivalent to the `inverse square law'. However the new theory of gravity also leads to the decay of orbits, and on the grounds of dimensional analysis we would expect comparable predictions. It is also usually argued that the Global Positioning System (GPS) demonstrated the efficacy of General Relativity. However as shown in [13] the new spatial-flow formalism of gravity also explains this system, and indeed gives a physical insight into the processes involved. In particular the relativistic speed and `gravitational red-shift' effects now acquire a unified explanation.

## Discussion

The experimental evidence from at least seven observations of absolute linear motion, some using 2$^{nd}$ order $(v/c)^2$ Michelson interferometers and some 1$^{st}$ order $v/c$ coaxial cable experiments, all showed that absolute linear motion is detectable, and indeed has been so ever since the 1887 Michelson-Morley experiment. Even Michelson and Morley reported a speed of 8km/s using the Newtonian theory for the instrument, but which becomes $\geq v_p = 300$ km/s when the Fitzgerald-Lorentz dynamical contraction effect *and* the refractive index effect are both taken into account. It then follows that vacuum interferometer experiments will fail to detect that absolute motion, as is the case. Reports from such `dud' experiments are always eagerly published by mainstream physics journals, as yet another confirmation of the Einstein postulates, but which merely promote even further ongoing confusion. These experiments are merely testing the Lorentz contraction effect, and so far no violations have been seen. We also understand that the various relativistic effects are caused by the absolute motion of systems through space, an idea that goes back to Lorentz. Elsewhere [2,3,7] we have shown that both the Galilean and Lorentz transformations have a role in describing mappings of data between observers in relative motion, but that they apply to different forms of the data. So absolute motion is a necessary part of the explanation of relativistic effects, and indeed the Lorentz transformation and symmetry are consistent with absolute motion, contrary to current beliefs within physics. On the contrary the Einstein postulates and their apparent link to these relativistic effects have always been understood to imply that absolute motion is incompatible with these relativistic effects. It was then always erroneously argued that the various observations of absolute motion over more than 100 years must have been flawed, since the relativistic effects had been well confirmed in numerous experiments.

So the Einstein postulates have had an enormously negative influence on the development of physics, and it could be argued that they have resulted essentially in a 100-year period of stagnation of physics, despite many exciting and valid developments, but even these will require a review of their deeper foundations, particularly in the cases of Electromagnetism, Relativistic Quantum Field Theory and Cosmology.

A major effect of the Einstein postulates was the development of a relativistic theory of gravity that was constrained to agree with Newtonian gravity in the non-relativistic limit. But indications that Newtonian gravity was flawed have been growing for over 50 years, as evidenced by the numerous so-called `gravitational anomalies', namely observations of gravitational effects incompatible with Newtonian gravity. The most well known of these is the `dark matter' effect, namely that spiral galaxies appear to require at least 10x the observed matter content in order to explain the high rotation speeds of stars and gas clouds in the outer regions. We now see that this effect is not caused by any form of matter, but rather by a non-relativistic but non-Newtonian aspect to gravity. As well the Greenland bore hole g anomaly data has revealed that the dimensionless constant that determines the magnitude of this spatial self-interaction dynamics is non other than the fine structure constant. The detection of absolute motion implies that space has some structure, for it is motion through that structure which is known as `absolute motion', and which causes relativistic effects. The phenomena of gravity are described by two gravitational constant, $G$ and $\alpha$, and it is the small size of $\alpha$ that determines the asymptotic form of the orbital rotation speeds in spiral galaxies. As well it is $\alpha$ that determines the magnitude of the black hole masses at the centres of globular clusters, and the Hubble Space Telescope data from M15 and G1 confirms that $M_{BH} = \alpha M_{GC}/2$, where $M_{BH}$ is the black hole mass, and $M_{GC}$ is the total mass of the globular cluster, in agreement with (8), see [6]. These are minimal black holes where the in-flow, mandated by the matter in the globular cluster, has a singularity. There the space is effectively destroyed. On the other hand primordial black holes are not mandated by matter, and are residual phenomena from the big bang. These primordial black holes form the kernels of spiral galaxies, and explain why these galaxies formed so quickly in the history of the universe. The properties of black holes in the new theory of gravity are determined by the value of $\alpha$, and not by $G$, which determines solely the interaction between matter and space. As well the new gravity theory has rotating black hole solutions, which supersede the Kerr-metric rotating black holes of GR. It needs to be emphasised that (8) is a classical equation for a velocity field, with the fine structure constant $\alpha$ appearing as a fundamental dimensionless constant playing now a critical role in spatial self-interaction dynamics. Nevertheless it suggests that the spatial dynamics is quantum dynamical at a deep level, and that the effects of that quantum dynamics manifests at a much more accessible scale than previously thought, where such quantum gravity effects were previously speculated to occur at the Planck length, time and mass scales.

The detection of absolute motion and the failure of Newtonian gravity together imply that General Relativity is not a valid theory of gravity; and that it is necessary to develop a new theory. This has now been achieved, and the essential task of checking that theory against experiment and observation has now explained all the known effects that GR was supposed to have explained, but most significantly, has also explained the numerous `anomalies' where GR was in manifest disagreement with the experimental or observational data. In particular a component of the flow past the earth towards the sun has been extracted from the analysis of the yearly variations of the Miller data [2,3,4].

The putative successes of the Einstein postulates lead to the Minkowski-Einstein spacetime ontology that has dominated the mindset of physicists for 100 years. Spacetime was mandated by the misunderstanding that absolute motion had not been observed, and indeed that it was incompatible with the established relativistic effects. Of course it was always possible to have chosen one foliation of the spacetime construct as the actual one separating the geometrical model of time from the geometrical model of space, but that never happened, and that possibility became one of the banned concepts of physics.

We are now in the position of understanding that space is a different phenomenon from time, that they are not fused into some spacetime amalgam, and that the spacetime ontolgy has been one of the greatest blunders in physics. This must not be misunderstood to imply that the numerous uses of a *mathematical* spacetime, particularly in Quantum Field Theory, were invalid.

There the mathematical spacetime formalism encodes dynamical effects of absolute motion. What is invalid is the assertion that such a mathematical spacetime is a physical entity.

We may now ask, for the first time in essentially 100 years, about the nature of space. It apparently has `structure' as evidenced by the fact that motion through it is detectable by various experimental techniques, and that its self-interaction is determined in part by the fine structure constant. As argued elsewhere [2,3,7] one interpretation is that space is a quantum system undergoing classicalisation, and at a deep substratum level has the characteristics of a quantum foam; such an ongoing `quantum collapse' would explain why the gravitational field $g$ is instantaneous from (5) and (8). This quantum foam is in differential motion, and the inhomogeneities and time dependencies of this motion cause accelerations which we know of as gravity. This motion is not motion of something through a geometrical space, but an ongoing restructuring of that quantum foam. One theory for this quantum foam arises in an *information-theoretic* formulation of reality known as *Process Physics*, in which all of reality is relational or internal information, so that in particular space itself is an information pattern, and one implication there is that the quantum-foam system undergoes exponential growth in a process of self-organisation, once the size of the quantum-foam system dominates over the matter part of the universe. This effect then appears to explain what is known as the `dark energy' effect, although of course it is not an energy at all, just as `dark matter' is not a form of matter. As well within this *Process Physics* we see a possible explanation for quantum matter, namely as informational topological defects embedded in the spatial quantum foam. That gives the first insights into an explanation for the necessity of quantum behaviour and also classicalisation. This first unification of matter and space also gives an insight into the origin of inertia, namely that matter effectively propagates through space at a constant velocity, unless subject to an applied force [2,3].

Another curious aspect to the ongoing confusion in physics about the fundamental meaning of motion is the spurious claim by Einstein and others that absolute rotational motion is meaningful, but that absolute linear motion is not, despite the fact that the latter is indistinguishable from the former in the limit of large radius orbits by matter. The problem here is fundamental and critical to our whole comprehension of reality. What determines whether matter is in motion or not? What is the inherent meaning of motion? Einstein of course considered at one stage that this could be explained by Mach's Principle, namely that absolute rotation effects were caused by rotation relative to distant matter in the universe. This whole debate goes back to Galileo's postulate that the earth rotated `absolutely' about its own axis, as well as `absolutely' about the sun. In his time this postulate was fiercely debated, and indeed he did not have any experimental or observational proof, though it clearly had simplicity. It was only in 1851 that Foucault, using his famous precessing pendulum experiment, finally demonstrated the absolute rotation of the earth about its own axis. But that discovery still left a real explanation missing. Of course it was Miller who finally clinched all the experimental evidence for this fundamental problem, although Michelson and Morley could and should have done so. Miller detected in 1925/26 the daily rotation of the earth about its own axis, as shown in Fig.4, and as well the rotation of the earth about the sun (using that effect to calibrate the interferometer instrument), and also the linear motion of the solar system (in a direction almost perpendicular to the plane of the ecliptic, which raises some interesting dynamical explanations for this). As well a recent reanalysis of his data, possible now that we have a relativistic theory for the operation of the interferometer that also takes account of the effects of the gas through which the light passes, has revealed a spatial in-flow past the earth towards the sun, as well as wave effects. We now understand, for the first time since Galileo's original propositions, the intrinsic meaning of motion; it is motion relative to the substructure of space, that space is a complex processing dynamical system, apparently a quantum-like information system, with the informational `patterns' in differential motion.

This quantum-foam spatial system invites comparison with the much older concept of the 'aether', but it differs in that the aether was usually considered to be some form of matter residing within a geometrical space, which is not the case here with the quantum foam theory of space; for here the geometrical description of space is merely a coarse grained description. Nevertheless it would be uncharitable not to acknowledge that the quantum-foam system is a modern version and indeed a return to the aether concept, albeit a banned concept.

Physics is a science. This means that it must be based on (i) experiments that test its theories, and (ii) that its theories and reports of the analyses of experimental outcomes must be freely reported to the physics community. Regrettably, and much to its detriment, this has ceased to be the case for physics. Physics has been in an era of extreme censorship for a considerable time; Miller was attacked for his major discovery of absolute linear motion in the 1920's, while DeWitte was never permitted to report to physicists the data from his beautiful 1991 coaxial cable experiment. Amazingly these experimenters were unknown to each other, yet their data was is in perfect agreement, for by different techniques they were detecting the same phenomenon, namely the absolute linear motion of the earth through space. All discussions of the experimental detections of absolute motion over the last 100 years are now banned from the mainstream physics publications. But using modern vacuum resonant cavity interferometer technology, and with a gas placed in the cavities, these devices could be used to perform superb high-precision experimental detections of absolute motion, and at least one such experiment is planned. As well the Miller and DeWitte data shows the presence of a wave phenomenon different to the waves argued to arise within GR theory, but which have not been detected, despite enormous costly efforts. It is now time to separate the genuine relativistic effects and their numerous manifestations from the flawed Einstein postulates, and to finally realise that they are caused by absolute motion of systems through a complex quantum system, which we know of as space. As for General Relativity it turns out to have been a major blunder. Nevertheless there is much evidence that a new theory of gravity has emerged, and this is to be exposed to critical analysis, and experimental and observational study. However the new theory of gravity has not yet been extended to a cosmologically compact growing closed space; such a space is implied by the deeper *Process Physics*.

The experimental evidence is, and that evidence dates from 1887, that the speed of light $c$ is the speed relative to a physically existing space. When measuring the speed of light an observer must take account of the effects of absolute motion on his measuring rod and clock otherwise, spuriously, the speed of light will *appear* to such an observer to be $c$, see [2,3]. The spacetime ontology is based on not understanding these real physical effects, despite Lorentz and others proposing them in the 19$^{th}$ century. Indeed the introduction of the spacetime construct is reminiscent of Ptolemy's epicycles; there the artefact of epicycles was finally abandoned when it was realised that they were spurious effects from not taking account of the fact that the observer was taking observations from an earth in orbit about the sun. Observers must always strive to be aware of any spurious effects introduced by a poorly understood measurement procedure, otherwise that spurious effect gets misinterpreted as a fundamental aspect of reality, as happened for both Ptolemy and Einstein.

For more detailed discussion of the new physics see the papers available at [19]. Support from Professor Igor Bray, Professor Warren Lawrence and Dr Lance McCarthy is gratefully acknowledged.

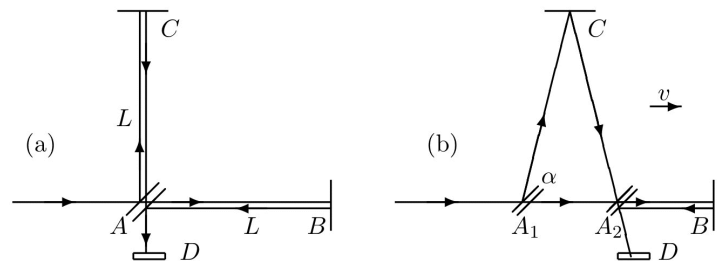

Fig.1 Michelson interferometer with beam splitter at A and mirrors at B and C. Here arms have equal length $L$. Fringe pattern is detected at D. (a) shows light paths when at rest, (b) shows light paths when in motion in direction parallel to arm AB.

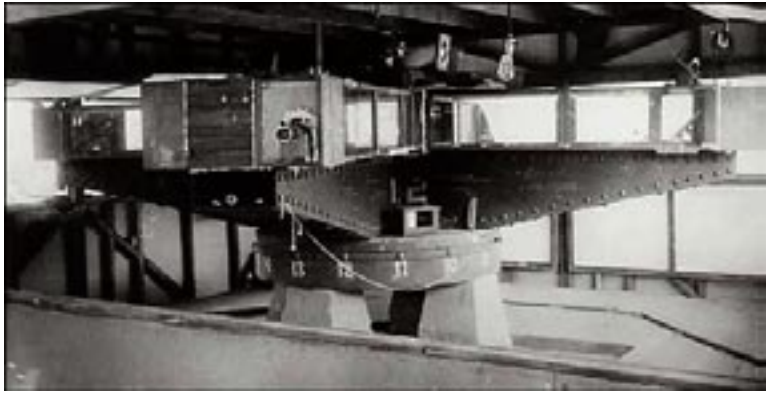

Fig.2 The Miller interferometer, at Mt. Wilson, had an effective arm length of $L$=32m, achieved by multiple reflections. The steel arms weighed 1200 kilograms and floated in a tank of 275 kilograms of mercury. Fringe shift readings were taken every $22.5^0$, as shown by the markers.

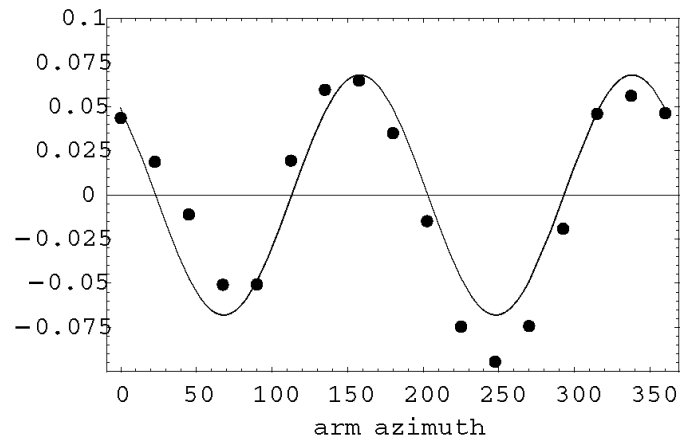

Fig.3 Typical Miller rotation induced fringe shifts taken every $22.5^0$, in fractions of a wavelength $\Delta\lambda/\lambda$, vs. azimuth $\theta$ (deg), measured clockwise from North, from a 20-turn average at Cleveland Sept. 29, 1929 16:24 UT; 11:29 Sidereal Time. This shows the quality of the fringe data that Miller obtained. The curve is the best fit to the form in (2), and gives $\psi = 158^0$, or $\psi = 22^0$ measured from South, and a projected speed of $v_P$=351 km/s.

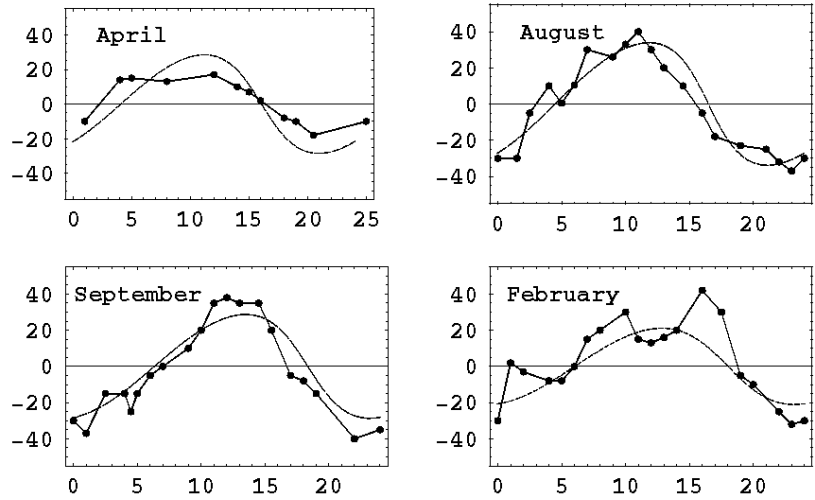

Fig.4 Azimuth ψ (deg), measured from south, from the Miller [5] data, plotted against local sidereal time. Plots cross the local meridian at approximately 5hr and 17hr. The monthly changes arise from the orbital motion of the earth about the sun. Miller used that effect to determine the value of k in (2), and which is now in agreement with the refractive index theory for the value of $k$, see [2,3,4]. Best fit curves from theory [2,4].

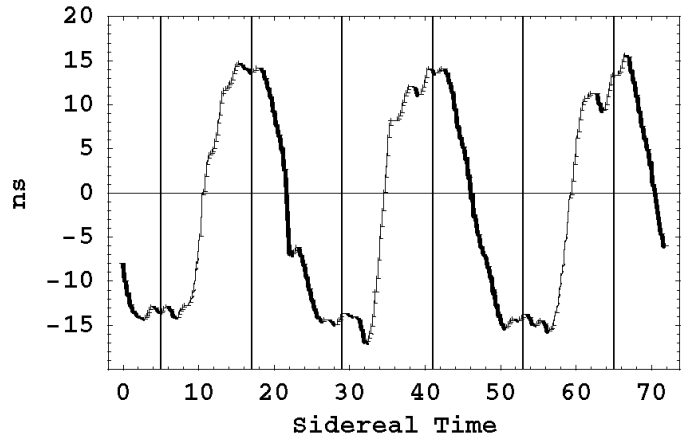

Fig.5 DeWitte 1991 RF travel-time variations, in ns, in a 1.5km NS orientated coaxial cable, measured with caesium atomic clocks, over three days and plotted against local sidereal time, showing that at approximately 5hr and 17hr the effect is largest. This remarkable agreement with the Miller interferometer experiment shows that the detection of absolute motion is one of the great

suppressed discoveries in physics. At least six other interferometer or coaxial cable experiments are consistent with this observation [2-4].

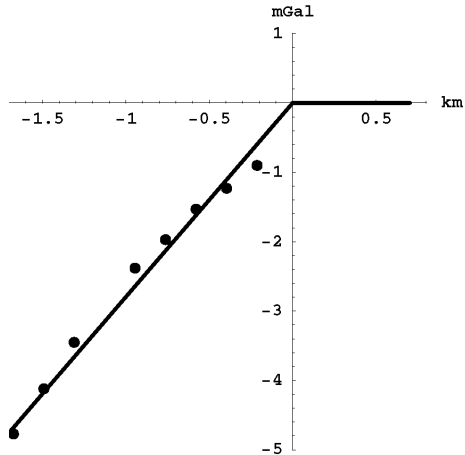

Fig.6 Gravity residuals from the Greenland bore hole anomaly data [17] measured in mGal (1mGal= $10^{-3} cm/s^2$). These are the differences in g ($\approx 9.8 m/s^2$) between the measured g and that predicted by the Newtonian theory. According to (8) this difference only manifests within the earth, see [6], and so permits the value of $\alpha$ to be determined. The data shows that $\alpha$ is the fine structure constant, to within experimental error. This small value for $\alpha$ explains why the spiral galaxy rotation speed plots are so flat, and also explains the observed black holes masses at the centre of globular clusters.

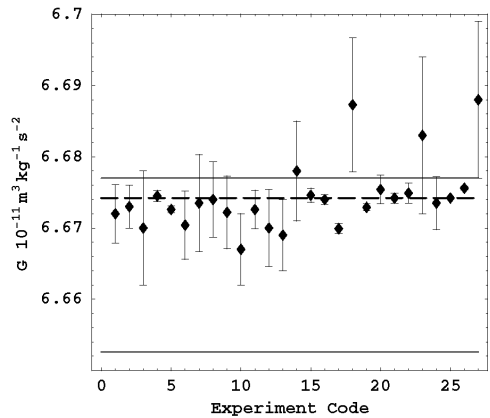

Fig.7 Results of precision measurements of G over the last sixty years in which the Newtonian theory of gravity was used to analyse the data. Shows that a systematic effect is missing from the Newtonian theory, of fractional size. $\approx \alpha/4$ For this reason G is the least accurately known fundamental constant. The upper horizontal line shows the value of G from an ocean Airy measurement [18], while the dashed line shows the current CODATA value. The lower horizontal line shows the value of G after correcting for the `dark matter' effect. Results imply that Cavendish laboratory experiments can measure $\alpha$.

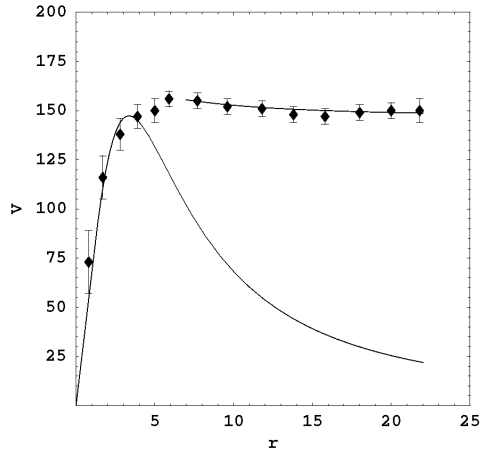

Fig.8 Data shows the non-Keplerian rotation-speed curve for the spiral galaxy NGC3198 in km/s plotted against radius in kpc/h. Lower curve is the rotation curve from the Newtonian theory or from General Relativity for an exponential disk, which decreases asymptotically like $1/\sqrt{r}$. This discrepancy is the origin of the `dark matter' story. The upper curve shows the asymptotic form from (13), with the decrease determined by the small value of α. This asymptotic form is caused by the primordial black holes at the centres of spiral galaxies, and which play a critical role in their formation. The spiral structure is caused by the rapid in-fall to these primordial black holes.

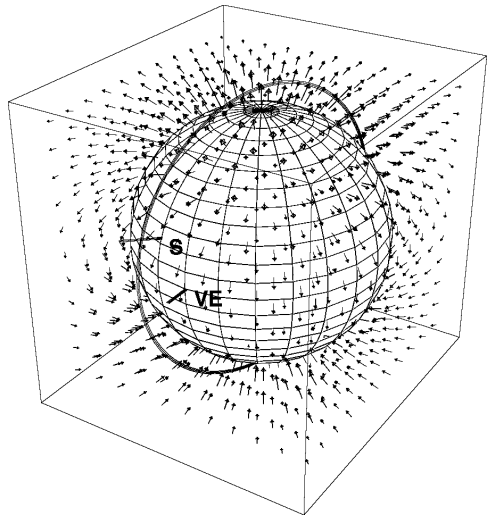

Fig.9 Shows the smaller component of the vorticity generated by the rotation of the earth, known as the Lense-Thirring effect. This has the form of a dipole field. In General Relativity only the earth-rotation induced vorticity is permitted, where it is known as a gravitomagnetic effect. Vorticity is a local rotation of space relative to more distant space. This rotation can be detected by observing the precession of a gyroscope, whose spin direction S remains fixed in the local space. VE is the vernal equinox. The Gravity Probe B satellite experiment [14,15] is designed to detect these precessions.

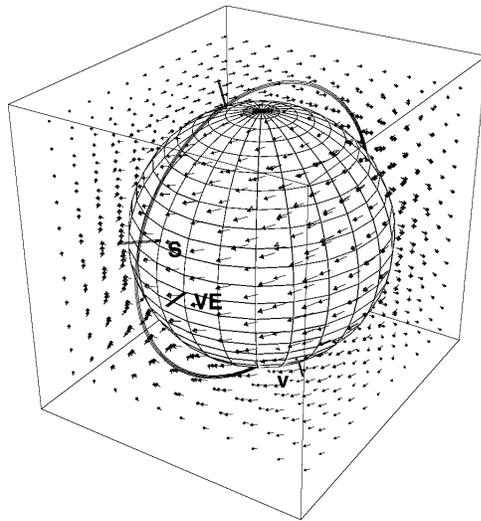

Fig.10 Shows the earth with absolute linear velocity V in the direction discovered by Miller [5] in 1925/26. This motion causes the vorticity field component shown by the vector field lines. A much smaller vorticity is generated by the rotation of the earth. Vorticity is a local rotation of space relative to more distant space. This rotation can be detected by observing the precession of a gyroscope, whose spin direction S remains fixed in the local space. VE is the vernal equinox. The Gravity Probe B satellite experiment [14,15] is designed to detect these precessions.

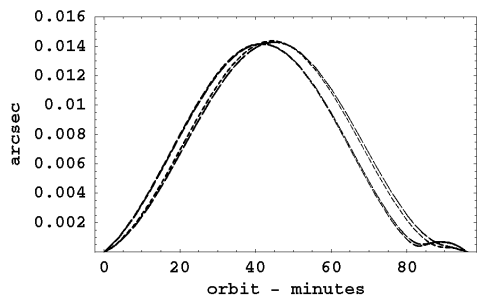

Fig.11 Shows predicted spin precession angle for the Gravity Probe B satellite experiment over one orbit, with the orbit shown in Fig.10, caused by the vorticity field arising from the absolute linear motion of the earth [11,12]. Plots are for four different months. This particular precession is not cumulative, compared to the precession from the earth-rotation induced vorticity component. The GP-B experiment is optimised to detect this smaller cumulative spin precession. General Relativity has earth-rotation induced vorticity but not the absolute linear motion induced vorticity.